\documentclass[]{spie}  

\usepackage{amsmath,amsfonts,amssymb}
\usepackage{graphicx}
\usepackage[colorlinks=true, allcolors=blue]{hyperref}
\usepackage{adjustbox}
\usepackage{xcolor}
\usepackage{bm}
\usepackage{multirow}
\usepackage{booktabs}
\usepackage[final]{microtype}
\usepackage{wrapfig}
\usepackage{ifthen}
\usepackage{amsmath}
\usepackage[ruled,vlined,linesnumbered]{algorithm2e}

\SetCommentSty{mycommfont}

\newboolean{showAppendix}
\setboolean{showAppendix}{false} 

\DeclareMathOperator*{\argmin}{arg\,min\xspace}

\newcommand{\MPS}{MPS$^{95}$\xspace}
\newcommand{\para}[1]{\noindent\textbf{#1}\quad}
\newcommand{\tagx}{\ensuremath{\cos \left(\gamma G_x \tau x\right)}\xspace}
\newcommand{\Lsim}{\ensuremath{\mathcal{L}_\mathrm{sim}}\xspace}
\newcommand{\Lsmooth}{\ensuremath{\mathcal{L}_\mathrm{smooth}}\xspace}

\title{Is Registering Raw Tagged-MR Enough for Strain Estimation~in~the Era of Deep Learning?}

\author[a]{Zhangxing Bian}
\author[b]{Ahmed Alshareef\,}
\author[a]{Shuwen Wei}
\author[a]{Junyu Chen}
\author[a]{Yuli Wang}
\author[c]{\\Jonghye Woo}
\author[d]{Dzung~L.~Pham}
\author[e]{Jiachen Zhuo}
\author[a]{Aaron Carass}
\author[a]{Jerry~L.~Prince}

\affil[a]{Johns Hopkins University, Baltimore MD, US}
\affil[b]{University of South Carolina, Columbia SC, US}
\affil[c]{Massachusetts General Hospital and Harvard Medical School, Boston MA, US}
\affil[d]{Uniformed Services University of the Health Sciences, Bethesda MD, US}
\affil[e]{University of Maryland School of Medicine, Baltimore~MD,~US}

\authorinfo{Further author information: (Send correspondence to Zhangxing Bian: zbian4@jhu.edu)}

\begin{document} 
\maketitle

\begin{abstract}
	Magnetic Resonance Imaging with tagging~(tMRI) has long been utilized for quantifying tissue motion and strain during deformation. However, a phenomenon known as tag fading, a gradual decrease in tag visibility over time, often complicates post-processing. The first contribution of this study is to model tag fading by considering the \textit{interplay} between $T_1$ relaxation and the repeated application of radio frequency (RF) pulses during serial imaging sequences. This is a factor that has been overlooked in prior research on tMRI post-processing.
	Further, we have observed an emerging trend of utilizing \textit{raw} tagged MRI within a deep learning-based~(DL) registration framework for motion estimation. In this work, we evaluate and analyze the impact of commonly used image similarity objectives in training DL registrations on \textit{raw} tMRI. This is then compared with the Harmonic Phase-based approach, a traditional approach which is claimed to be robust to tag fading.  Our findings, derived from both simulated images and an actual phantom scan, reveal the limitations of various similarity losses in raw tMRI and emphasize caution in registration tasks where image intensity changes over time.
\end{abstract}

\keywords{MR tagging, deep learning registration, image similarity, strain estimation}

\section{INTRODUCTION}
\label{sec:intro}

Tagged Magnetic Resonance Imaging (tMRI)~ \cite{axel1989heart, axel1989mr} is a technique that has been employed for numerous years to vizualize and quantify the intricate movements of tissues undergoing deformation~\cite{mcveigh1998imaging,parthasarathy2007measuring,knutsen2014improved}. 
This technique is used in many applications including assessing myocardial function~\cite{kolipaka2005relationship, ibrahim2011myocardial}, studying motion during speech and swallowing~\cite{parthasarathy2007measuring, xing2019atlas, gomez2020analysis, shao2023analysis}, and measuring brain motion after low-acceleration impacts~\cite{knutsen2014improved}.
MR tagging temporarily magnetizes tissue with a spatially modulated periodic pattern, creating transient tags in the image sequence that move with the tissue, thus capturing motion and strain information. 

Processing tMRI must contend with the phenomenon of tag fading, characterized by a gradual decrease in tag visibility over time. The cause of tag fading is traditionally ascribed to $T_1$ relaxation. 
In the present study, we delve deeper into the modelling of tag fading, considering the \textit{interplay} between $T_1$ relaxation and the repeated application of radio frequency~(RF) pulses during serial image acquisition. 
This understanding is important because the fading pattern of tags substantially influences subsequent estimations of motion and strain.
Algorithms like optical flow and image registration are often used to track tag motion across time frames~\cite{osman2000visualizing, PVIRA, bian2023midl, ye2021deeptag, ye2023sequencemorph, bian2023momentamorph}. However, the fading of tags disrupts the premise of brightness constancy, which in turn hampers tracking and yields less precise estimates of tissue motion. 

The Harmonic Phase (HARP) method is a widely recognized approach for processing tMRI and is reported to exhibit robustness against tag fading~\cite{osman1999cardiac,osman2000imaging,osman2000visualizing}. HARP applies a bandpass filter on harmonic peaks in $k$-space to obtain the harmonic phase, which is assumed to be an innate, spatially-varying property of the material being imaged. HARP-based tracking approaches~\cite{PVIRA, bian2023midl, yu2023new} employ harmonic phase images in the registration framework for motion tracking, and thus should not be influenced by the brightness changes induced by tag fading.

Recent advances in deep learning-based registration techniques have given rise to a new approach that directly employs \textit{unsupervised} registration algorithms on \textit{raw} tagged-MRI, using normalized cross-correlation~(NCC) as a similarity objective~\cite{ye2021deeptag, ye2023sequencemorph}. The underlying assumption is that the preservation of brightness constancy may not be imperative, provided that the model is equipped with appropriate image similarity objectives, such as NCC. This notion is similar to the premise of inter-modality registration tasks, where the intensity profile of the image pair does not necessarily exhibit a clear correlation or relationship.

This inspires us to assess the effectiveness of deep registration methods with diverse similarity objectives when dealing with tag fading. It also raises the question of whether harmonic phase-based techniques are becoming outdated in the deep learning~(DL) era.
In this study, we systematically compare the HARP-based and raw tag-based approaches using simulated tagged data undergoing a known motion and a real motionless phantom; therefore, in both cases the motion is known. 
Our analysis provides insights on how tag fading influences motion and strain estimation, and how different similarity objectives impact results.  
On a broader scale, this work reveals caveats and limitations of various similarity losses in registration tasks where the image intensity changes over time. Such scenarios can be found in pre- and post-intervention images and in the evolution of an injected contrast agent with patient motion.

The contributions of this work are summarized as follows:
\textbf{(1)}~We develop a mathematical model to capture the phenomenon of tag fading, factoring in the \textit{interplay} of $T_1$ relaxation and transition to the steady state.
\textbf{(2)}~We evaluate and analyze the impact of widely-used similarity losses in training DL registration on tMRI. Simulated images and a real phantom scan are used to quantify the error of motion and strain, which are essential outputs in biomechanical studies. 

\section{Method}

\begin{figure}[!tb]
	\centering
	\includegraphics[width= 0.9\linewidth]{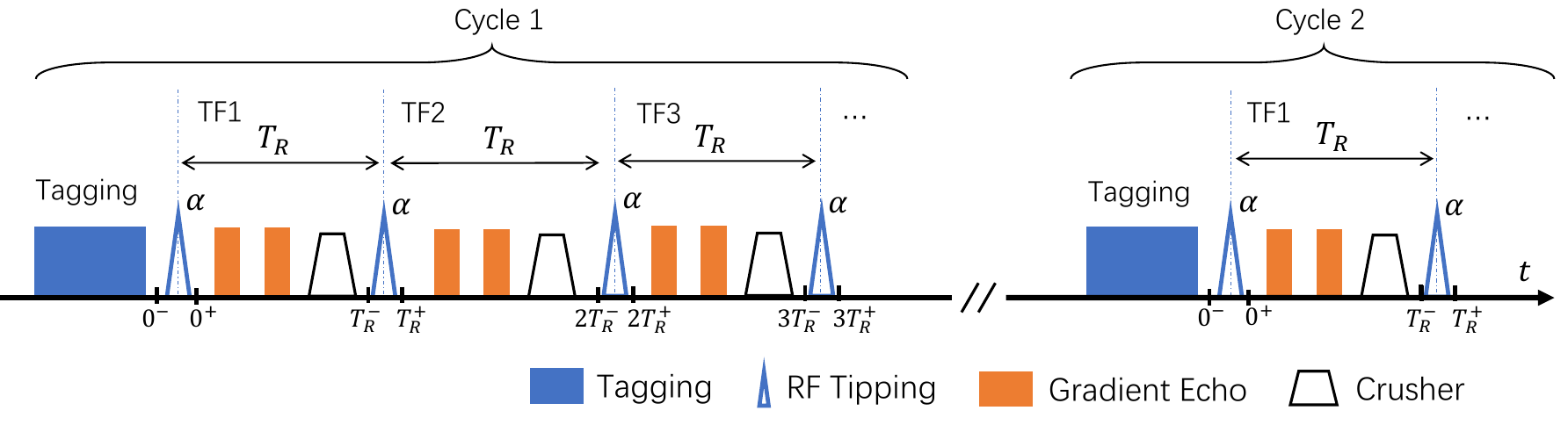}
	\caption{tMRI imaging sequence. ``TF'' stands for time-frame.}
	\label{fig:sequence}
	\vspace{-1em}
\end{figure}

For our analysis, we have selected the classic 1:1~SPAMM~\cite{park1996analysis} tagging sequence, which is shown in Fig.~\ref{fig:sequence}. The illustration demonstrates the process, wherein each tagging step is followed by a series of imaging sequences. During each $T_R$ interval, the spin system is tipped by $\alpha$ and multiple line segments in $k$-space are captured using gradient echoes. To prepare for the next $T_R$ interval, a crusher gradient is applied to dephase the magnetization in the transverse plane. The ``tagging-imaging" cycle is repeated until sufficient $k$-space coverage is achieved. 
We denote  the magnetization along the $z$-axis as $M_z$. 
We ignore the subtlety that the lines in $k$-space are acquired with varying amplitudes during $T_2^*$ decay. Consequently, the amplitude of taglines at the $n^{\text{th}}$ repetition time $T_R$ are proportional to $M_z(x,nT_R^{-})$, which can be iteratively represented in terms of $M_z(x,(n-1)T_R^{-})$, given an initial condition, as
\begin{align}
	M_z(x, n T_R^{-}) &= M_z\left(x,(n-1) T_R^{+}\right) \cdot e^{-\frac{T_R}{T_1}} + M_0\left(1-e^{-\frac{T_R}{T_1}}\right) \label{eq:1-1}\\
	&= M_z\left(x,(n-1) T_R^{-}\right) \cos(\alpha) \cdot e^{-\frac{T_R}{T_1}} + M_0\left(1-e^{-\frac{T_R}{T_1}}\right)  \label{eq:1-2} \\
	M_z\left(x, 0^{-}\right) &= M_0 \tagx \label{eq:1-3}\,.
\end{align}
Equation~\ref{eq:1-1} applies the $T_1$-relaxation model. Equation~\ref{eq:1-2} illustrates the application of the RF tipping with the tip angle $\alpha$. Equation~\ref{eq:1-3} represents the cosine pattern of magnetization that is introduced into the tissue immediately after the tagging process. In these equations, \( x \) denotes spatial location, \( M_0 \) the equilibrium magnetization magnitude, \( \gamma \) the gyromagnetic ratio, and \( G_x \) the x-direction magnetic gradient.
With this tag-fading model, we can simulate 1:1 SPAMM tMRI sequences.

\begin{figure}
	\centering
	\includegraphics[width= 0.7\linewidth]{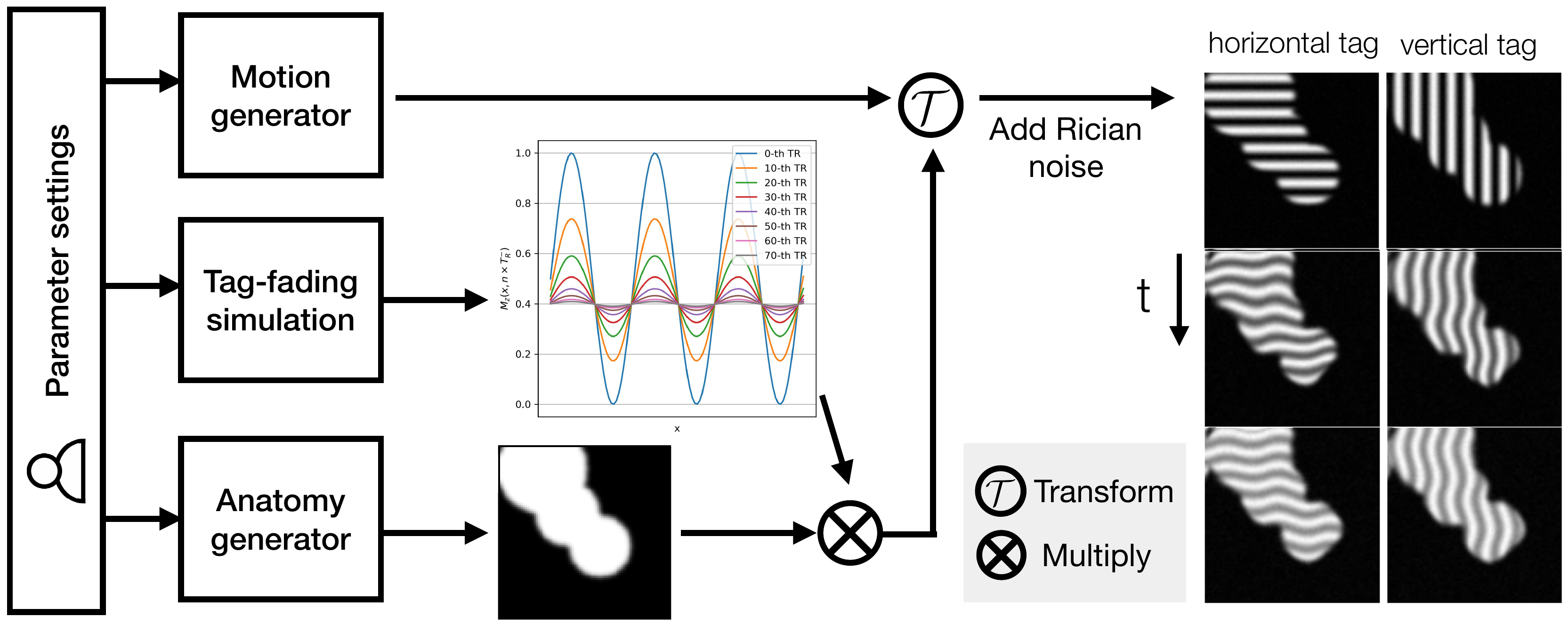}
	\caption{Simulation pipeline.}
	\label{fig:simulation}
	\vspace{-1em}
\end{figure}

\para{Simulation} In this section, we simulate a dataset for DL model training and evaluation. Shown in Fig.~\ref{fig:simulation}, the simulation pipeline is composed of: the motion generator, the tag-fading simulation, and the anatomy generator.
The motion generator creates an elastic deformation field that mimics tissue motion, while the tag-fading simulation manages the amplitude of tags at each time point, governed by Eqns.~\ref{eq:1-1} through~\ref{eq:1-3}. The anatomy generator produces random disks whose union represents the anatomy being imaged.
The process of tagging tissue is modeled as the multiplication of anatomy with tag patterns. The images are then warped by the generated deformation field to simulate deformed tag lines. Rician noise (with $\sigma=0.02$) is subsequently added. \looseness=-1

\para{Registration framework}
Typically, DL-based registration frameworks aim to learn a function $\bm{\phi} = g_{\theta}(F, M)$, where $\bm{\phi}$ is the transformation used to align the fixed image $F$ and the moving image $M$. The function $g$ is parameterized by the learnable parameters $\theta$. These parameters are learned by optimizing a generalized objective:
\begin{equation}
	\hat{\theta} =  \argmin_{\theta} \Lsim(F, M \circ g_{\theta}(F, M)) + \lambda \Lsmooth(g_{\theta}(F, M))\,.
\end{equation}
\Lsim encourages image similarity between the fixed, $F$, and warped moving, $M$, images. 
\Lsmooth imposes a smoothness constraint on the transformation and $\lambda$ weights the relative impact of two terms.  

It is not feasible to evaluate all DL-based registration architectures. Instead, we selected the popular convolutional neural network~(CNN)-based registration architecture, VoxelMorph~\cite{voxelmorph2019}, as our network backbone.
We considered various similarity losses, including mean square error ($\mathtt{MSE}$), normalized cross correlation~\cite{avants2008symmetric}($\mathtt{NCC}$), mutual information~\cite{viola1997alignment} ($\mathtt{MI}$), structural similarity index~\cite{wang2004image} ($\mathtt{SSIM}$), normalized gradient fields~\cite{haber2006intensity} ($\mathtt{NGF}$), and modality independent neighborhood descriptor~\cite{heinrich2012mind,heinrich2013towards} ($\mathtt{MIND}$).  These choices encompass the most widely-used similarity measurements for both mono- and multi-modal registration. For a comprehensive discussion on the characteristics of the various similarity losses, we refer readers to the review~\cite{chen2023survey}.
We encourage the spatial smoothness of the displacement $\bm{u}$, with the smoothness loss $\mathcal{L}_\mathrm{smooth} = \sum_{\bm{x}} \| \nabla \bm{u}(\bm{x}) \|^2$.

\section{Experiments}
\paragraph{Materials \& Training Details} 
We generated a dataset comprising 1,000 ``\textit{movies}" through simulation (Fig.~\ref{fig:simulation}), where each movie consists of 40 time frames, each of size \(96 \times 96\). Each movie exhibits a unique anatomy, which subsequently undergoes random elastic deformation. During the generation process, the values for \( T_1 \) and \( T_R \) are randomly sampled from the intervals \([800, 1000]\)ms and \([15, 25]\)ms, respectively. The RF tip angle is set to \( \alpha = 15^\circ \). The tag period is determined as 10\% of the image dimension, i.e., \( 9.6 \) pixels.
We split the data samples into training, validation, and testing datasets using a 6:2:2 ratio, respectively. In addition, we acquired a real tMRI sequence consisting of 78 time frames by scanning a static cylindrical gel phantom filled with Sylgard 527 in a 1:1 ratio, with material properties that emulate human brain tissue. Note that this phantom has no motion and the ground truth is zero deformation. This real phantom data is exclusively reserved for testing.

To facilitate a fair comparison, we independently searched for the optimal hyperparameters for each choice of \Lsim using a Bayesian optimizer~\cite{ozaki2020multiobjective}, which has proven to be a more effective strategy than grid search.
For instance, we searched for the bin size and Parzen window size for $\mathtt{MI}$, the window size for  $\mathtt{NCC}$, and the illumination, contrast, and structural factors for $\mathtt{SSIM}$. The weight $\lambda$ for \Lsmooth was also searched. 
Given that the anatomical masks of the simulated images are available, we employed the Dice coefficient between the fixed and warped moving masks as the metric during the hyperparameter search. 
An alternative option is to employ End-Point Error (EPE) as the metric, given the availability of ground truth for synthetic deformation. However, in practical scenarios, pixel-wise ground truth is typically unavailable, making EPE an impractical performance indicator for the selection of hyperparameters. 
During both training and evaluation, we used the first frame as the fixed image, and sampled moving images from all the subsequent frames. 
We modified the initial convolutional layer's input channels to accommodate both raw vertical and horizontal-tagged images, as well as sinusoidal-transformed HARP (sHARP) images~\cite{bian2023midl}, which is a variant of traditional HARP. The sHARP effectively eliminates discontinuities in phase images, thereby facilitating seamless end-to-end training.

\para{Evaluation metrics}
Given a sequence of tagged images, we create registration pairs by selecting the first image as the fixed image and pairing it with each subsequent image as the moving image. Registration accuracy is assessed using the End Point Error~(EPE), which measures the magnitude of deviation between the estimated and ground truth displacement vector fields.
We evaluate the impact on strain computation by calculating the $95^{\text{th}}$ percentile error of the maximum principle strain~(\MPS) for each static test pair, which quantifies erroneous strain measurements. For instance, a \MPS value of 0.07 signifies that 5\% of the tissue undergoes length stretching of at least 7\%. Evaluation on each registration pair yields a EPE and \MPS.

\para{Results} 
We evaluate and compare the performance of the raw-based and HARP-based approaches under various selections of \Lsim, as well as SyN~\cite{avants2008symmetric} with various similarity metrics, all of which are summarized in Table~\ref{tab:quanti}.
Generally, the HARP-based input outperforms the raw tMRI input in terms of  EPE and \MPS. Specifically, the ``raw + MSE'' method attempts to match the absolute profile of the faded tag pattern, which often leads to the creation of illusory motion. NCC performs relatively better with the raw tMRI since it captures the correlation of a local intensity profile (within a window), making it less sensitive to changes in absolute intensity. However, it still generates errors that are statistically significantly higher than those produced when using HARP-based images as input.
The overall best performance is achieved when using the HARP-based input trained with NCC loss. The harmonic phase, being a material property, spans the range $[-\pi, \pi)$. Ideally, a perfect estimation would be achievable through matching the \textit{absolute} phase value under MSE. However, our observations indicate that NCC slightly outperforms MSE in both EPE and \MPS.

\definecolor{deepred}{HTML}{f07167}
\definecolor{lightred}{HTML}{d68c45}
\definecolor{deepgreen}{HTML}{2c6e49}
\definecolor{lightgreen}{HTML}{1dd3b0}
\definecolor{lightblue}{HTML}{add8e6}
\definecolor{blue}{HTML}{0000ff}
\newcommand{\maxval}[1]{\textcolor{deepred}{#1}}
\newcommand{\maxvals}[1]{{#1}}
\newcommand{\minval}[1]{\bm{\textcolor{lightgreen}{#1}}}
\newcommand{\minvall}[1]{\textcolor{blue}{#1}}

\begin{table}[!tb]
	\centering
	\caption{Quantitative Results. The \minval{best}, \minvall{second best}, and \maxval{worst} performing method in
each column are highlighted. Wilcoxon signed-rank tests were conducted on pairs (labeled by brackets on the right) for all four columns, with * indicating statistical significance with $p < 0.01$ (with Bonferroni correction).}
	\includegraphics[width= 0.9\linewidth]{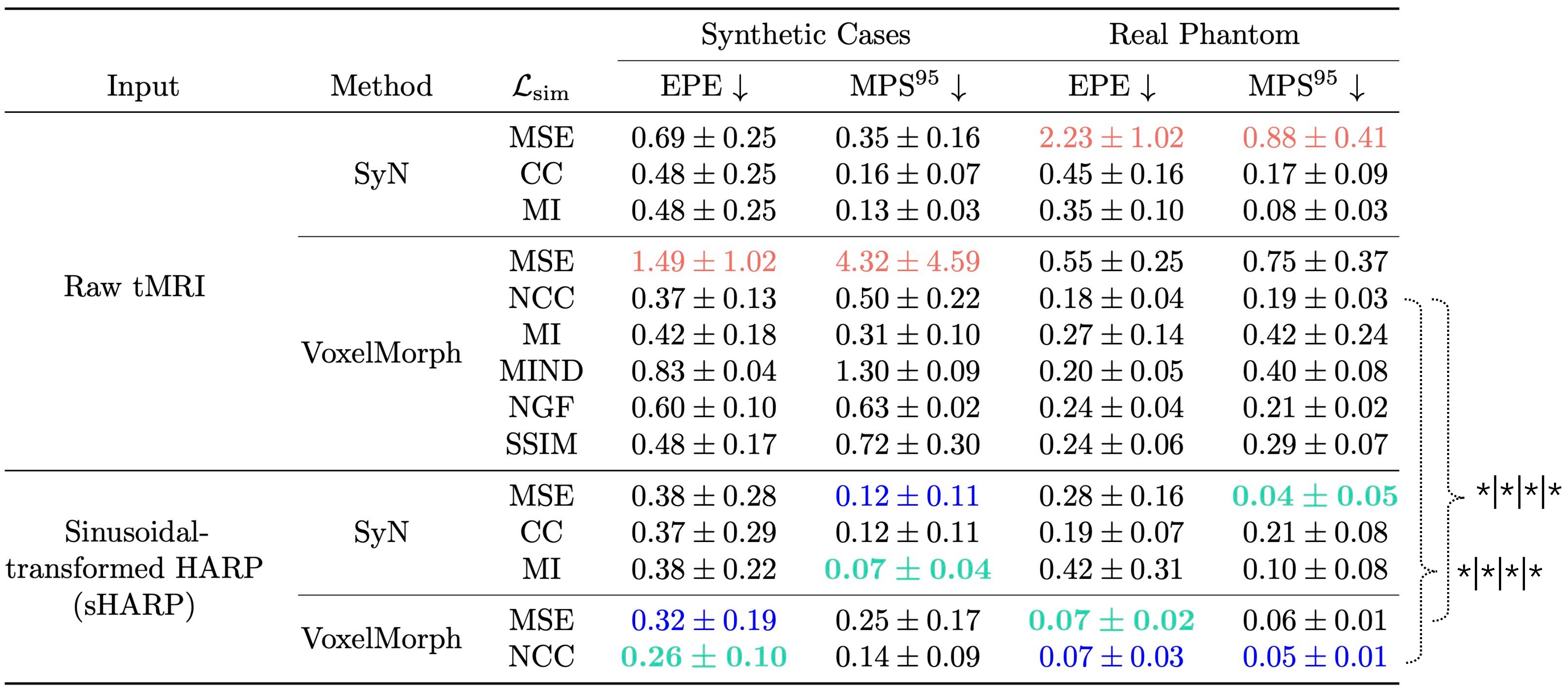}
	\label{tab:quanti}
\end{table}

\begin{figure}[!tb]
	\centering
	\includegraphics[width= \linewidth]{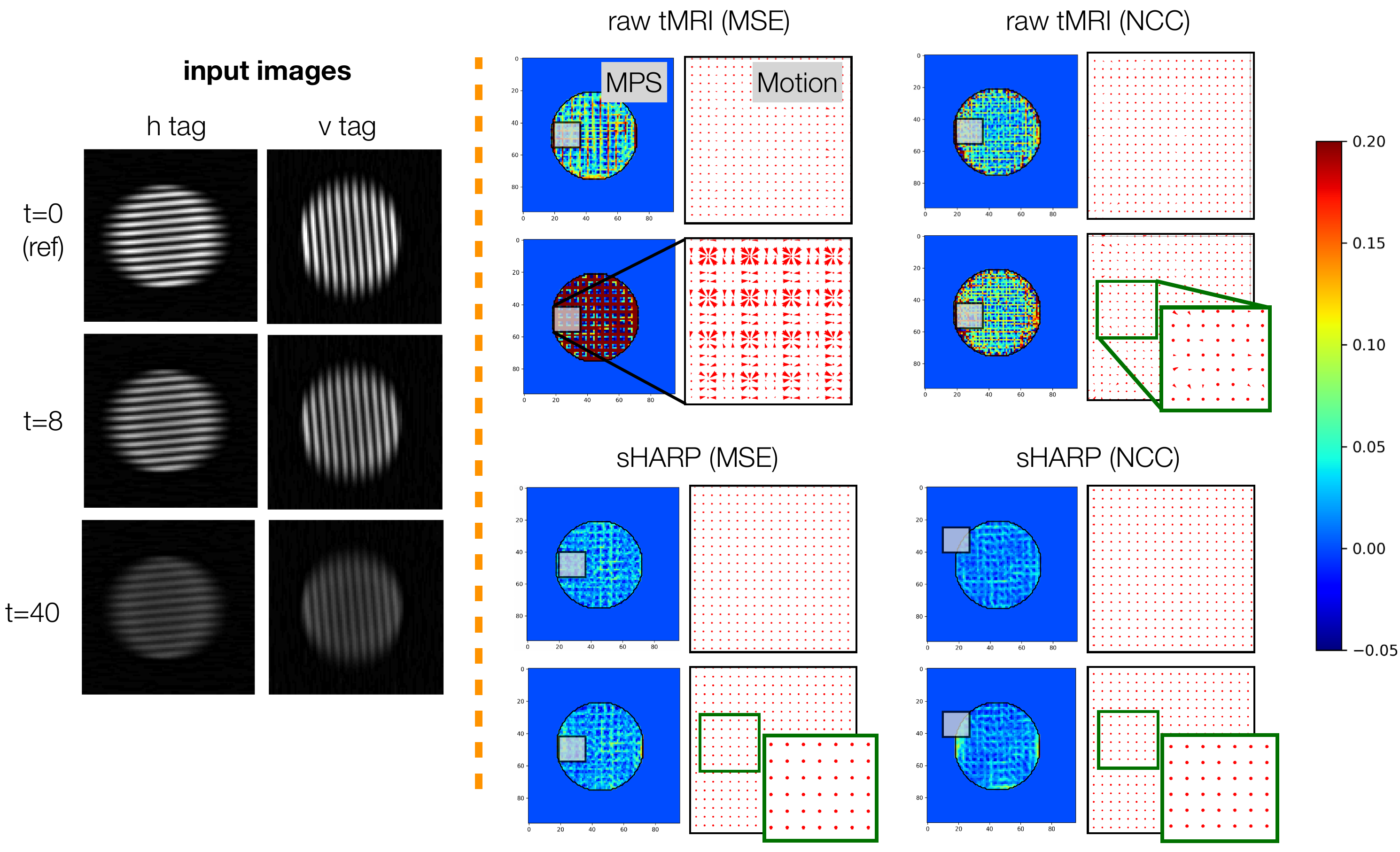}
	\caption{Qualitative results on real motionless data. Left panel shows horizontal (h-tag) and vertical (v-tag) tagging for three timeframes. Right panel shows the color-coded \MPS map and zoomed-in motion field of four methods. The first and second row show the registration results of ref-to-8$^{\text{th}}$ and ref-to-40$^{\text{th}}$ frame.\looseness=-1}
	\label{fig:quali}
\end{figure}

\section{Conclusion}

In this study, we constructed a mathematical model of MRI tagging to depict the phenomenon of tag fading, accounting for the interplay between $T_1$ relaxation and the progression toward steady-state during repeated imaging pulses, an aspect overlooked in previous research. We also take note of the recent trend in DL that uses raw tMRI as an input for motion or strain estimation. We perform a comparative analysis using Harmonic Phase (HARP) images versus raw tMRI as inputs for DL-based registration techniques. Our findings indicate that HARP-based input methods are less prone to generate false motion or strain when tag fading is present.

For future work, we intend to delve into deep similarity metric learning~\cite{czolbe2021semantic, ronchetti2023disa}, which have shown promise in inter-modality image registration and are presumed to be robust against the challenges posed by tag fading. Although our current research is limited to SPAMM sequences, the potential of Complementary-SPAMM~(CSPAMM) sequences to enhance tag contrast and mitigate tag fading cannot be ignored, despite their longer acquisition times. Current study of tag-fading dynamics could inform enhancements to the SPAMM sequence, aiming to achieve CSPAMM-level contrast while retaining rapid acquisition times.

\section*{Acknowledgments}
This work was supported in part by the NIH through the National
Institute of Neurological Disorders and Stroke~(NINDS) grant
U01-NS112120~(PI: P.V.~Bayly) and the National Institute on Deafness
and Other Communication Disorders~(NIDCD) grant R01-DC018511~(PI: J.
Woo). The opinions and assertions expressed herein are those of the
authors and do not reflect the official policy or position of the
Uniformed Services University of the Health Sciences or the Department
of Defense.

\small
\bibliography{report} 
\bibliographystyle{spiebib} 

\end{document}